\documentclass[%
 reprint,
superscriptaddress,
 amsmath,amssymb,
 aps,
prd,
]{revtex4-2}

\usepackage{graphicx}
\usepackage{dcolumn}
\usepackage{bm}
\usepackage{hyperref}
\usepackage{booktabs}
\usepackage{multirow}
\usepackage{xcolor}
\usepackage{comment}
\usepackage[normalem]{ulem}
\usepackage{cleveref}

\usepackage{titlesec}
\titleformat{\paragraph}
  {\normalfont\normalsize \it}{\theparagraph}{1em}{}
\titlespacing*{\paragraph}
  {0pt}{3.25ex plus 1ex minus .2ex}{1.5ex plus .2ex}
\usepackage{orcidlink}

\begin{document}


\title{Astrophysical Constraints for Bosonic Dark Matter in and Around Quark Stars
}

\author{J\'ulia Tartera Barber\`a~\orcidlink{0009-0004-2767-458X}}
 \email{jtartera@ice.csic.es}
\affiliation{Institute of Space Sciences (ICE, CSIC), Campus UAB, Carrer de Can Magrans, 08193 Barcelona, Spain}

\author{Sarah Louisa Pitz~\orcidlink{0009-0007-4169-4298}}
\email{pitz@itp.uni-frankfurt.de}
\affiliation{Institut f\"ur Theoretische Physik, Goethe Universit\"at, 
  Max-von-Laue-Str.~1, 60438 Frankfurt am Main, Germany}

\author{J\"urgen Schaffner-Bielich~\orcidlink{0000-0002-0079-6841}}
\email{schaffner@astro.uni-frankfurt.de}
\affiliation{Institut f\"ur Theoretische Physik, Goethe Universit\"at, 
  Max-von-Laue-Str.~1, 60438 Frankfurt am Main, Germany}

\author{Laura Tolos~\orcidlink{0000-0003-2304-7496}}
 \email{tolos@ice.csic.es}
\affiliation{Institute of Space Sciences (ICE, CSIC), Campus UAB, Carrer de Can Magrans, 08193 Barcelona, Spain}
\affiliation{Institut d'Estudis Espacials de Catalunya (IEEC), 08860 Castelldefels (Barcelona), Spain}

\date{\today}

\begin{abstract}
We investigate the structure, stability, and mass-radius relations of compact stars composed of quark matter within the non-ideal MIT Bag Model mixed with self-interacting bosonic dark matter with a $\phi^n$ self-interaction potential. By solving the coupled two-fluid Tolman-Oppenheimer-Volkoff equations for boson masses between 200 MeV and 1 GeV  and self-interaction strengths with $n=4$ and $n=40$, we show that strong self-interactions and higher boson masses enhance the stability of compact star configurations at high central pressures, though they reduce maximum overall masses. Ultra-compact solutions with $C \ge 1/3$ capable of mimicking black holes are significantly restricted compared to hadronic models, emerging only for strong self-interactions and dark matter fractions exceeding 90\%. Depending on the parameters, the compact stars form either extended dark matter halos or compact cores: stronger self-interactions and higher boson masses suppress the radial extension to favor compact cores of dark matter. Matching our models against observational constraints highlights PSR J1614$-$2230 and XTE J1814$-$338 as potential window of opportunity for dark matter detection. Notably, XTE J1814$-$338 could confirm a dark sector, given its requirement for light boson masses ($m_b \approx 200\text{--}300\text{ MeV}$) and strongly DM-dominated configurations.

\end{abstract}

\maketitle

\section{\label{sec:introduction} Introduction}

As stars reach the end of their evolutionary life, they leave behind dense remnants known as compact objects. The nature of the remnant depends primarily on the mass of the progenitor star. Stars below 8$M_\odot$ leave behind white dwarfs, while more massive stars become neutron stars or black holes. Since nuclear fusion no longer takes place inside white dwarfs and neutron stars, they are supported against further gravitational collapse by quantum effects such as degeneracy pressure and strong nuclear interactions.

Neutron stars can have masses ranging from theoretical minimum values of roughly $1.2 M_\odot$ \cite{minimalmassNS1,minimalmassNS2}, up to theoretical maximum masses of about $3 M_\odot$ \cite{Lattimer:2021emm}. The most massive confirmed pulsars have measured masses close to $2.35 M_\odot$ \cite{MostMassiveNS}.  Recent measurements by the NICER mission have constrained the radii of pulsars such as PSR J0030+0451  \cite{Miller_2019, Riley_2019,Vinciguerra:2023qxq,Kini:2026rjx}, PSR J0740+6620 \cite{Riley2021,Miller_2021} and PSR J0614-3329  \cite{Mauviard:2025dmd,Miller:2026vpr}, with values of $\sim$ 10-12 km. Confining such extreme masses within these radii makes neutron stars some of the most compact objects in the Universe.

Traditionally, these dense remnants have been modelled as stars composed of ordinary baryonic matter (see e.g.\ \cite{Schaffner-Bielich_2020} for an introduction). However, several recent astronomical observations remain difficult to explain within this conventional picture. Examples include objects with unexpectedly low masses and small radii, such as HESS J1731-347 and XTE J1814-338 \cite{anomaloussmallmass, anomalouscompactnessstar}, as well as gravitational-wave events like GW190814 and GW230529 \cite{GW190814, GW23}, which involve components with masses close to or above the expected theoretical limits for neutron stars with nuclear matter in their interior.

These observational tensions have motivated the exploration of alternative descriptions of compact stars, which serve as unique laboratories for matter under conditions irreproducible on Earth. Despite decades of research, the exact internal composition of these stars remains uncertain. At sufficiently high densities, matter may undergo a phase transition in which nucleons dissolve into their constituent quarks. As a result, the core, or possibly the entire compact star, could be composed of deconfined quark matter (QM), forming what is known as a strange quark star \cite{PhysRevD.30.272, bodmer1971collapsed, 1970PThPh..44..291I}.

At the same time, a wide range of astrophysical and cosmological observations indicate the presence of dark matter (DM), a non-luminous form of matter that interacts gravitationally but has not yet been directly detected. Evidence for DM arises from galactic rotation curves, gravitational lensing, galaxy cluster dynamics, and measurements of the cosmic microwave background, which together indicate that DM constitutes approximately 85\% of the matter content of the Universe \cite{Bertone_2005, Planck2018, Betoule_2014, Massey2007, CosmologicalParameters2004}. These observations suggest that ordinary matter alone may not fully describe the composition of compact stars, and that DM could play a role in their internal structure.

The nature of DM remains one of the major open questions in modern physics. Traditionally, the standard cold DM paradigm has been based on collisionless candidates such as Weakly Interacting Massive Particles (WIMPs). Although collisionless cold DM successfully reproduces the large-scale structure of the Universe, it faces tensions with several small-scale astrophysical observations \cite{Tulin_2018}. Consequently, self-interacting DM models have attracted considerable attention as a possible solution to these discrepancies \cite{Cassing_2023}.
In recent years, compact stars have become an important astrophysical probe for studying DM properties \cite{Goldman_1989,Bertone_2008, Brayeur_2012, Bertone_2018, de_Lavallaz_2010}. Due to their strong gravitational fields, compact stars can capture DM from their surroundings, allowing it to accumulate and potentially alter the stellar structure \cite{Kouvaris_2010, Brito_2015,Kouvaris_2008_cooling}. Moreover, specific models allow for the existence of purely dark objects such as boson stars, self-gravitating spheres that could have emerged via standard structure formation in the early Universe \cite{Chang:2018bgx,Pitz_BosonStars}. Thus, possible scenarios are dark stars accreting ordinary matter \cite{Meliani:2016rfe,Betancourt_Kamenetskaia_2023}, binary mergers between a standard compact remnant and a dark star \cite{Diedrichs_2023}, or the decay of standard model particles into the dark sector \cite{Baym_2018, Motta_2018,Husain_2022, Colpi_1986}. Consequently, the study of compact stars containing both ordinary matter and DM has become an active field of research. Early studies primarily focused on admixed neutron stars, investigating how a DM component modifies their structure, stability, and gravitational-wave signatures \cite{Leung_2011, Das2020_MonNot, Kain_2021,pitz2024_om_dm,Barbat_2024, Rather:2026aed, Thakur:2026pes}, but also exploring the properties of DM admixed quark stars \cite{Mukhopadhyay2016, panotopoulos2025, Sen_2022,Ferreira_2023,Lenzi_2023}.
Motivated by these studies, this work explores the interplay between self-interacting bosonic DM and QM in compact objects. Using a two-fluid model in which both components interact only through gravity, this work investigates how DM affects the structure, stability, and observable properties of quark stars and whether the admixture can yield to ultra-compact configurations. This study extends previous work of two of us \cite{pitz2024_om_dm}, which considered strongly self-interacting bosonic DM with ordinary baryonic matter to the case of QM for compact stars. The ultimate goal is to further investigate the possible role of DM in compact stars and its implications for the interpretation of current and future astrophysical observations. In our work we restrict the DM particle mass $m_b$ to a range of a few MeV to GeV, since it was shown by \cite{Colpi_1986} that the maximum mass of the star scales with $\propto 1 / m_b^2$ which would results in either too large or too small masses of the compact star if the particle mass is outside the specified range.
The paper is organized as follows. In Sec.~\ref{sec:TheoreticalFramework} we present the theoretical framework, introducing the Tolman-Oppenheimer-Volkoff equations \ref{subsec:TwoFluidTOV}, the stability analysis of the TOV solutions \ref{sec:StabAnalysis}, the equation of state for QM and DM \ref{sec:EoS} and the corresponding speed of sound and fluid stiffness \ref{sec:speed_of_sound}. Our results in Sec.~\ref{sec:results} are focused on the stability and compactness \ref{sec:results_stability_compactness}, mass-radius relations \ref{subsec:Mass_radii} and observational constraints \ref{subsec:Observational_constraints}. We finalize our paper with the conclusions in Sec.~\ref{sec:conclusions}.

\section{\label{sec:TheoreticalFramework} Theoretical Framework}

\subsection{\label{subsec:TwoFluidTOV} Tolman-Oppenheimer-Volkoff equations}

The macroscopic properties of compact stars, such as their mass and radius, are determined by the Tolman-Oppenheimer-Volkoff (TOV) equations, which describe the conditions for hydrostatic equilibrium accounting for the effects of general relativity. This work focuses on compact stars composed of QM and DM. To study the hydrostatic equilibrium of admixed compact stars, the standard TOV formalism is extended following the work of Sandin et al. \cite{Sandin_2009}. At present, all observational evidence for DM is based on its gravitational effects. Although some theories allow for additional interactions between DM and ordinary matter, these interactions are expected to be extremely weak compared to the gravitational fields inside compact stars \cite{Bertone_2005}. It is therefore common to model the system as two fluids interacting only through gravity. The resulting TOV equations using natural units ($\hbar = c = 1$), are
\begin{align}
    \frac{dP_{QM}(r)}{dr} &= - [\varepsilon_{QM}(r) + P_{QM}(r)] \alpha'(r)\\
    \frac{dm_{QM}(r)}{dr} &= 4\pi r^2\varepsilon_{QM}(r), \\
    \frac{dP_{DM}(r)}{dr} &= - [\varepsilon_{DM}(r) + P_{DM}(r)] \alpha'(r)\\
    \frac{dm_{DM}(r)}{dr} &= 4\pi r^2\varepsilon_{DM}(r),
\end{align}
where $P_i$ and $\varepsilon_i$ are the pressure and the energy density of each species and the derivative of the metric function $\alpha'(r)$ is 
\begin{equation*}\label{eq:alpha}
    \alpha' = \frac{Gm(r)}{r^2} \left( 1 + \frac{4\pi r^3 P(r)}{m(r)} \right) \left( 1 - \frac{2Gm(r)}{r} \right)^{-1},
\end{equation*}
with the total pressure and enclosed mass given by the sum of the contributions from both fluids ($P(r) = P_{QM}(r) + P_{DM}(r)$ and $m(r) = m_{QM}(r) + m_{DM}(r)$). 
The spatial extent of each fluid is determined by the radial coordinate at which its internal pressure vanishes. Thus, the individual radii defining the boundary of each component are $R_{QM}$ (where $P_{QM}(R_{QM}) = 0$) and $R_{DM}$ (where $P_{DM}(R_{DM}) = 0$). Consequently, the total radius of the compact star, $R$, is defined by the fluid that extends the farthest: $R = \max(R_{QM}, R_{DM})$. The total mass of the object is then given by the sum of the masses enclosed within their respective boundaries, $M = m_{QM}(R_{QM}) + m_{DM}(R_{DM})$.

For numerical calculations, it is convenient to rewrite the TOV equations in dimensionless form. Following the rescaling procedure described by Narain et al. \cite{Narain_2006}, we introduce a reference energy density scale $\varepsilon_0$ and define the dimensionless pressure and energy density as 
\begin{align}
    P' = \frac{P}{\varepsilon_0}, \quad \quad \varepsilon' = \frac{\varepsilon}{\varepsilon_0} ,
\end{align}
and the resulting transformations for the radial coordinate and mass are
\begin{align}
    r &= \frac{1}{\sqrt{G \varepsilon_0}} r', \\
    m &= \frac{1}{\sqrt{G^3 \varepsilon_0}} m'.
\end{align}

\subsection{Stability analysis}
\label{sec:StabAnalysis}

Compact stars are in hydrostatic equilibrium when the inward gravitational pull is cancelled out by the outward pressure gradient. Solving the TOV equations for a given central pressure $P_c$ therefore yields equilibrium configurations characterized by their mass $M$ and radius $R$. Moreover, the TOV equations can be supplemented with an additional equation for the particle number, allowing us to determine the total number of particles $N$ confined within the star. This value will be necessary for the stability analysis, as discussed later in this section and is obtained with
\begin{align}
    \frac{dN_{QM}(r)}{dr} &= 4\pi r^2 n_{QM}(r) \left( 1 - \frac{2Gm(r)}{r} \right)^{-1/2}, \\ \frac{dN_{DM}(r)}{dr} &= 4\pi r^2 n_{DM}(r) \left( 1 - \frac{2Gm(r)}{r} \right)^{-1/2}, \label{eq:TOV_particles}
\end{align}
where $m(r)$ is the total mass enclosed within radius $r$, $n_{QM}(r)$ and $n_{DM}(r)$ are the local number densities of the QM and DM fluids, respectively. The total particle numbers for each component are then obtained by integrating these equations up to their respective boundaries, $N_{QM} = N_{QM}(R_{QM})$ and $N_{DM} = N_{DM}(R_{DM})$.
Since the particle number equation is integrated together with the TOV system, the particle number must be rescaled consistently as well. Introducing a reference number density scale $n_0$ such that $n' = n/n_0$, it becomes
\begin{align}
    N &= \frac{n_0}{\sqrt{(G \varepsilon_0)^3}} N'.
\end{align}

To determine whether an equilibrium configuration is stable, the star must be subjected to infinitesimal radial perturbations. Physically, this corresponds to slightly compressing the star and studying the competition between gravity and the restoring pressure force. The strength of this restoring force depends on the stiffness of matter. If the restoring pressure is sufficiently strong, the perturbation remains oscillatory and the configuration is stable.

Mathematically, radial perturbations are introduced through a harmonic displacement of the form $\delta r(r,t)=\beta(r)e^{-i\omega t}$. Linearizing the Einstein field equations then leads to a Sturm-Liouville eigenvalue problem \cite{Misner:1973prb, Kokkotas_2001}, whose solutions determine the normal oscillation modes of the star. These modes are ordered according to their squared frequencies, $\omega_0^2 < \omega_1^2 < \omega_2^2 < \dots$

For a stable configuration, all squared frequencies must be positive ($\omega_i^2 > 0$), implying oscillatory perturbations. If any mode satisfies $\omega_i^2 < 0$, the frequency becomes imaginary and the perturbation grows exponentially. Since the modes are strictly ordered, the transition from stability to instability is entirely dictated by the fundamental mode $\omega_0^2$.

At the critical point where $\omega_0^2 = 0$, the perturbation becomes static and represents a displacement to an infinitesimally adjacent equilibrium state ($P_c \to P_c + \delta P_c$). During such perturbations, the total particle number remains conserved. In the single-fluid case, variations in the total mass are directly related to variations in the particle number \cite{Harrison1965_Gravitation}. Consequently, the transition point satisfies both $dN = 0$ and $dM = 0$.

The stability onset occurs at the turning point of the equilibrium sequence with respect to the central pressure \cite{zeldovich}, where
\begin{equation}
    \frac{dM}{dP_c}=0,
\qquad
\frac{dN}{dP_c}=0.
\end{equation}
According to the turning-point criteria developed by Shapiro et al. \cite{shapiro}, changes in stability are related to the sign of $dR/dP_c$ at the extrema of the mass curve. When the stellar radius decreases at the turning point ($dR/dP_c < 0$), a mode with an even number of radial nodes changes stability ($\omega_0$, $\omega_2$, \dots). Conversely, if the radius increases ($dR/dP_c > 0$), a mode with an odd number of nodes changes stability ($\omega_1$, $\omega_3$, \dots).

For a two-fluid system, the one-dimensional sequence of equilibrium solutions becomes a two-dimensional parameter space described by $(P_{c}^{QM}, P_{c}^{DM})$. In this case, the single-fluid maximum-mass criterion is no longer sufficient. Here we follow the formalism introduced by Hippert et al. \cite{Hippert}. Because the QM and DM components interact only gravitationally, a radial perturbation causes the star to behave as a macroscopic system of coupled harmonic oscillators. Consequently, the solutions to the Sturm-Liouville problem are not independent motions of isolated fluids but coupled normal modes.
Unlike in the single-fluid case, the total mass changes with the perturbations. However, the total particle numbers of the QM and DM sectors remain independently conserved. Therefore, the stability boundary is determined by configurations for which infinitesimal variations of the central pressures leave both particle numbers unchanged simultaneously,
\begin{align} 
&\begin{pmatrix}
\delta N_{QM} \\ \delta N_{DM}
\end{pmatrix}=
\\
  &  \begin{pmatrix}
    \partial N_{QM}/ \partial P_{c}^{QM} & \partial N_{QM}/ \partial P_{c}^{DM} \\
    \partial N_{DM}/ \partial P_{c}^{QM} & \partial N_{DM}/ \partial P_{c}^{DM}
\end{pmatrix}
\begin{pmatrix} \delta P_{c}^{QM} \\ \delta P_{c}^{DM} \end{pmatrix}= \begin{pmatrix} 0 \\ 0 \end{pmatrix}.\nonumber
\end{align}

To determine the stability region, we mathematically decouple the system by diagonalizing this matrix. This transformation gives us a diagonal matrix containing two eigenvalues, $\kappa_A$ and $\kappa_B$
\begin{equation} 
\begin{pmatrix}
\delta N_A \\ \delta N_B
\end{pmatrix}=
    \begin{pmatrix}
    \kappa_A & 0 \\
    0 & \kappa_B
    \end{pmatrix}
\begin{pmatrix} \delta P_{c}^A \\ \delta P_{c}^B \end{pmatrix} = \begin{pmatrix} 0 \\ 0 \end{pmatrix}.
\end{equation}
For a small shift in central pressure ($\delta P_c \neq 0$) to trigger an instability without changing the total number of particles ($\delta N = 0$), the determinant of the Jacobian must be zero ($\det \mathcal{J} = \kappa_A \kappa_B = 0$). The stability boundary is therefore reached when one of the eigenvalues crosses zero. Extending the single-fluid criterion to two dimensions, a two-fluid configuration is stable only if both eigenvalues remain positive ($\kappa_A > 0$ and $\kappa_B > 0$).

Computationally, the stability analysis requires solving the two-fluid TOV equations on a two-dimensional grid of central pressures. At each grid point, the particle numbers are computed, the Jacobian matrix is evaluated numerically, and its eigenvalues are used to identify the stability boundary.

\subsection{Equation of state}
\label{sec:EoS}
The Equation of State (EoS) is the thermodynamic relation connecting pressure $P$ to energy density $\varepsilon$. To solve the TOV equations, it is necessary to specify an EoS for each species. 
\subsubsection{Quark matter}
\label{subsec:QM_EoS}

In order to describe the deconfined state of quark matter, we employ the  non-ideal bag model, as shown in Refs.~\cite{Fraga:2001id,Alford:2004pf}. This model includes an empirical correction to the MIT Bag model so as to incorporate the effects of gluon-mediated QCD interactions between the quarks in the Fermi sea. This is done by introducing an adimensional parameter $a_4$, coming from the thermodynamic potential to one loop order,  that is related to the strong coupling constant $\alpha_s$ as $a_4 = (1 - 2\alpha_s/\pi)$\cite{Alcock:1986hz,Haensel:1986qb}.  When $a_4$ = 1, the original MIT bag model is recovered \cite{MITBagModel}.

Within this model, the number density ($n_q$), energy density ($\varepsilon_q$) and pressure ($P_q$) for each quark flavour $q$ is given by
\begin{eqnarray}
&&n_q = a_4 N_c \frac{k_f^3}{3\pi^2},  \nonumber \\
&&\varepsilon_q = a_4\frac{N_c}{\pi^2}\int_0^{k_{F}}E_q k^2 dk , \nonumber \\
&& P_q = a_4\frac{N_c}{3\pi^2}\int_0^{k_{F}} \frac{k^4 dk}{E_q} , 
\end{eqnarray}
with $N_c$ the number of colors, $E_q$ the single-particle energy for each quark flavour and $k_{F}$ the associated Fermi momentum. The single particle energy is given by $E_q=\sqrt{m_q^2+k^2}$, where $m_q$ is the quark mass. We use the current quark masses $m_u = 2.16$ MeV, $m_d = 4.7$ MeV, $m_s = 93.5$ MeV \cite{ParticleDataGroup:2024cfk}. Heavier quark flavors are not considered since their masses are much larger than the typical chemical potentials expected inside compact stars. 

The total energy density and pressure of quark matter can be then calculated as

\begin{equation}
\varepsilon_{QM} = \sum_q \epsilon_q + B \quad \mbox{and} \quad P_{QM}=  \sum_q p_q - B, \label{eq:EoS}
\end{equation}
with $B$ the bag constant.

The dynamics of neutron stars is governed by weak interactions. Weak interactions continuously convert quark flavors through processes such as
\begin{align*}
     &u + d \leftrightarrow u + s, \quad u + e^- \leftrightarrow d + \nu_e, \quad d \leftrightarrow u + e^- + \bar{\nu}_e, \\ 
     \quad &u + e^- \leftrightarrow s + \nu_e, \quad s \leftrightarrow u + e^- + \bar{\nu}_e.
\end{align*}
 In an evolved neutron star, neutrinos freely escape from the star ($\mu_\nu = 0$). The conditions of chemical equilibrium imply
\begin{equation}
    \mu_d = \mu_u + \mu_e, \qquad \mu_d = \mu_s. \label{eq:chemical_equilibrium}
\end{equation}
In addition, the stellar matter must remain electrically neutral, requiring
\begin{equation}
    \frac{2}{3}n_u - \frac{1}{3}n_d - \frac{1}{3}n_s - n_e = 0. \label{eq:charge_neutrality}
\end{equation}
The total baryon number density of the system is simply the sum of the individual quark densities divided by three (since three quarks constitute one baryon):
\begin{equation}
    n_B = \frac{1}{3}(n_u + n_d + n_s). \label{eq:baryon_density}
\end{equation}

For a fixed baryon density $n_B$, the chemical potentials of the particle species are obtained by solving the chemical equilibrium and charge neutrality conditions (Eqs. \ref{eq:chemical_equilibrium}, \ref{eq:charge_neutrality}), which are known as $\beta$-equilibrium. Once the chemical potentials are known, the pressure and energy density of Eq.~(\ref{eq:EoS}) can be evaluated by using the Fermi momentum $k_F = \sqrt{\mu_q^2 - m_q^2}$.  The exact analytical expressions for $n_q$, $\varepsilon_q$, and $P_q$ for each massive fermion species are given by
\begin{align}
    n_q &= a_4 \frac{1}{\pi^2} k_F^3,  \nonumber \\
    \varepsilon_q &= a_4 \frac{3}{8\pi^2} \left[ k_F \mu_q \left(2k_F^2 + m_q^2\right) - m_q^4 \ln\left(\frac{k_F + \mu_q}{m_q}\right) \right], \nonumber \\
    P_q &= a_4 \frac{1}{8\pi^2} \left[ k_F \mu_q \left(2k_F^2 - 3m_q^2\right) + 3m_q^4 \ln\left(\frac{k_F + \mu_q}{m_q}\right) \right]. \label{eq:massive}
\end{align}
The total macroscopic energy density and pressure of the stellar fluid are computed by summing these individual finite-mass contributions and incorporating the bag constant as established in Eq.~\eqref{eq:EoS}, yielding the final EoS.

The viability of the model made of up, down and strange quarks (strange quark matter, SQM) depends on the choice of the bag constant $B$, which is constrained by the Bodmer-Witten-Terazawa hypothesis \cite{PhysRevD.30.272,bodmer1971collapsed}. We first redefine $a_4$ as $a_4 = 1- c$, as done in Ref.~\cite{Alford:2004pf}, so the original MIT bag model is recovered with $c = 0$. In Ref.~\cite{Lopes:2020btp} it was found that the stability window for SQM can be obtained for different values of $c$. More precisely, increasing $c$ from 0 to 0.3 displaces the minimum value of $B^{1/4}$ from 148 MeV to 135 MeV, whereas the maximum value of $B^{1/4}$ changes from 159 MeV to 145 MeV. Moreover, the maximum mass of neutron stars was computed in Ref.~\cite{Lopes:2020btp} for the different $c$ and $B^{1/4}$ correlated pairs. It was concluded that a maximum mass of 2.21 $M_{\odot}$ is obtained for the case of $c=0.3$ and $B^{1/4}=135$ MeV, thus fulfilling the observational constraints of 2$M_{\odot}$. Therefore, for the numerical implementations in this work, we fix the bag constant parameter to $c=0.3$ and $B^{1/4} = 135$ MeV.

\subsubsection{Self-interacting bosonic dark matter}
\label{sec:DM_EoS}

To model the DM sector, we follow the approach of Pitz et al. \cite{Pitz_BosonStars}, describing the DM fluid as a gas of massive bosons through a self-interacting complex scalar field $\phi$.

In the present work we model the interaction using a generalized potential of the form $V(|\phi|^2) = \frac{\lambda}{2^{n/2}} |\phi|^n$, where $\lambda$ is the coupling constant and the exponent $n$ controls the interaction regime.

The resulting EoS is given by:
\begin{equation}
    \varepsilon' = P'\frac{n+2}{n-2} + P'^{2/n} \frac{4}{\lambda'^{2/n} \left(\frac{n}{2}-1\right)^{2/n}},
\end{equation}
where $P'$ is the DM pressure and $\varepsilon'$ the energy density.

To further simplify this relation, we introduce a secondary scaling parameter $\alpha$, defining $\tilde{P} = P'/\alpha$ and $\tilde{\varepsilon} = \varepsilon'/\alpha$. By setting the dimensionless coupling constant to $\tilde{\lambda} = {2^{n+1}}/({n-2})$, the coefficient $\alpha=1 $ and the EoS becomes
\begin{equation}
    \tilde{\varepsilon} = \tilde{P}\frac{n+2}{n-2} + \tilde{P}^{2/n}.
    \label{eq:DM_Dimensionless_EoS}
\end{equation}
This EoS depends on two free parameters, the boson mass $m_b$ and the interaction exponent $n$, which control the properties of the DM fluid.

To compute the total DM particle fraction within the star, we require the local particle number density $\tilde{n}_{DM}$:

\begin{equation}
    \tilde{n}_{DM} = \tilde{P}^{2/n} \sqrt{ \frac{2n}{n-2} \tilde{P}^{1-2/n} + 1 }.
\end{equation}

Since this expression is independent of the coupling constant, the DM particle distribution can be computed directly alongside the hydrostatic pressure profile.

\begin{figure*}
    \centering
    \includegraphics[width=0.99\textwidth]{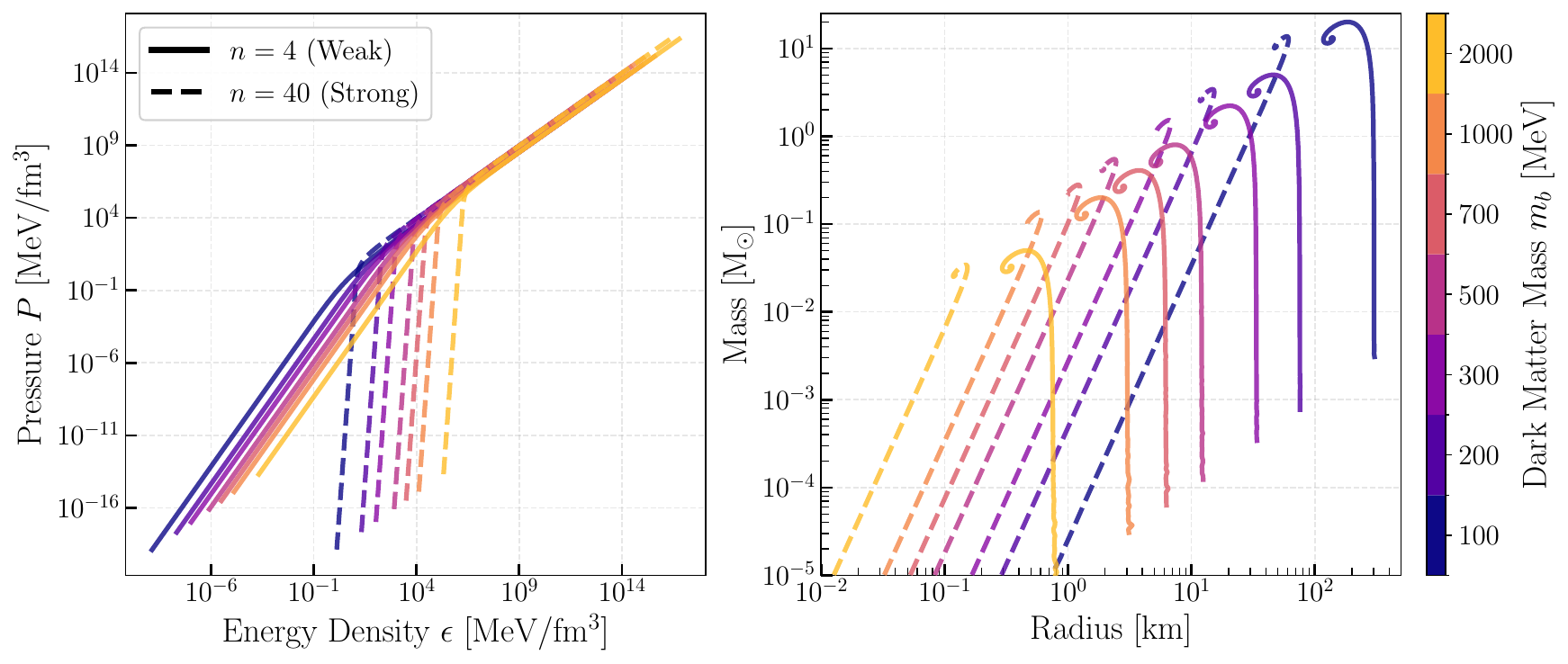} 
    \caption{\textbf{(a)} EoS and \textbf{(b)} the corresponding mass-radius relations for a single-fluid bosonic DM model varying the boson mass $m_b$ (color scale) and the self-interaction strength $n$ (dashed lines for $n=40$ and solid lines for $n=4$).}
    \label{fig:bdm_eos_mr_parametric}
\end{figure*}

\begin{table}[tbp]
    \centering
    \caption{Maximum mass configurations for purely bosonic DM stars across different values of the interaction parameter $n$ and boson mass $m_b$. The table details the maximum achievable mass $M_{\mathrm{max}}$, along with its corresponding central pressure $P_c$ and radius $R$.}
    \label{tab:dm_max_masses}
    \begin{tabular}{ccccc}
        \toprule
        $n$ & $m_b$ [MeV] & $M_{\mathrm{max}}$ [$M_\odot$] & $P_c( M_{\mathrm{max}})$ [MeV/fm$^3$] & $R(M_{\mathrm{max}})$ [km] \\
        \midrule
        4  & 200  & 5.01 & $1.10 \times 10^{1}$ & 46.44 \\
        4  & 300  & 2.22 & $5.58 \times 10^{1}$ & 20.64 \\
        4  & 500  & 0.80 & $4.30 \times 10^{2}$ & 7.43  \\
        4  & 1000 & 0.20 & $6.89 \times 10^{3}$ & 1.86  \\
        \addlinespace 
        40 & 200  & 3.43 & $3.55 \times 10^{2}$ & 14.84 \\
        40 & 300  & 1.52 & $1.80 \times 10^{3}$ & 6.60  \\
        40 & 500  & 0.55 & $1.39 \times 10^{4}$ & 2.37  \\
        40 & 1000 & 0.14 & $2.22 \times 10^{5}$ & 0.59  \\
        \bottomrule
    \end{tabular}
\end{table}

Figure \ref{fig:bdm_eos_mr_parametric} shows the EoS and the corresponding mass-radius relations for a purely bosonic DM fluid, illustrating the effect of the boson mass $m_b$ and the self-interaction strength $n$ on the stellar structure. Table \ref{tab:dm_max_masses} summarizes the maximum mass configurations for each parameter combination, including the corresponding central pressure and radius.

The parameter values are selected to explore the different regimes that will be discussed in Section \ref{sec:results}. The boson mass $m_b$ is varied between 100 MeV and 2 GeV, while the self-interaction strength is studied for two representative cases following Pitz et al. \cite{pitz2024_om_dm}: a weakly interacting regime with $n=4$ (solid lines) and a strongly interacting regime with $n=40$ (dashed lines).

The case $n=4$ corresponds to the standard $\phi^4$ interaction commonly considered in boson star models. In contrast, $n=40$ represents a stronger self-interaction that leads to the ultra-compact configurations, that are of interest for this work.

In the weakly interacting $n=4$ models, the EoS is softer than that of the strongly interacting $n=40$ models, which effectively means that the central pressures it can sustain are smaller (as quantitatively shown in Table \ref{tab:dm_max_masses}). This is typically associated with lower maximum masses. However, for $n=4$ the energy density smoothly approaches zero ($\varepsilon \propto P^{1/2}$) alongside the pressure. This gradual decline allows the fluid to remain gravitationally stable over vast distances, allowing radii to become $\sim 3$ times larger than those of their strongly interacting counterparts. Because the spherical volume scales as $V \propto R^3$, this extensive spatial distribution enables the configuration to continuously integrate mass across its massive outer layers. As a result, the weakly interacting models reach exceptionally high maximum masses despite requiring relatively modest central pressures.

Conversely, the strongly interacting $n=40$ EoS behaves similarly to a self-bound fluid, maintaining a high energy density even as the local pressure nearly vanishes. This produces a much sharper drop in the pressure profile and confines the star to smaller radii. As a result, despite being able to sustain much higher central pressures and exhibiting a nearly incompressible core, the total mass is limited by the smaller spatial extension of the star.

We now turn to the effect of the boson mass $m_b$. Changing the value of $m_b$ between 100 MeV and 2 GeV, we obtain a wide range of compact star masses and radii in Fig. \ref{fig:bdm_eos_mr_parametric}(b). Because both the maximum mass and the corresponding radius of the star scale identically as $\propto m_b^{-2}$, their ratio is conserved ($M_{\mathrm{max}} \propto R_{\mathrm{max}}$). Thus, on the log-log scale of this figure, this linear proportionality translates into a diagonal line along which the mass-radius curves are displaced as $m_b$ is varied.

Physically, increasing the boson mass (moving from the dark blue to the yellow curves) shifts the EoS toward higher energy densities for any given pressure. This scaling effectively softens the fluid, meaning it provides less outward pressure support to counteract gravity. Consequently, heavier bosons scale the curves down toward significantly smaller radii and lower maximum masses.

\subsection{Speed of sound and fluid stiffness}
\label{sec:speed_of_sound}
To compare the stiffness of the different EoSs introduced in this section, we analyse the speed of sound of each fluid as the squared speed of sound $c_s^2$ measures the response of the EoS to compression.

\begin{figure}
\centering
\includegraphics[width=\linewidth]{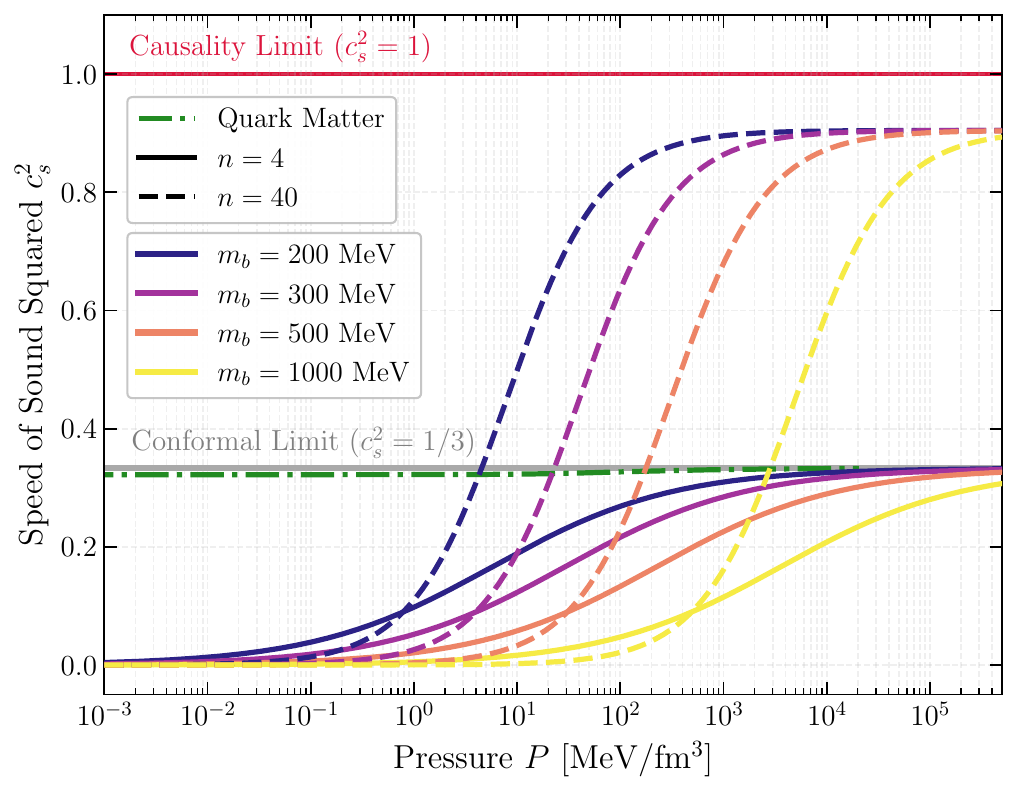}
\caption{Evolution of the squared speed of sound ($c_s^2$) as a function of pressure for the massive QM model and the bosonic DM models across a range of boson masses ($m_b$) and interaction strengths ($n=4, 40$). The conformal (green dashed-dotted line) and causality (red solid line) limits are indicated.}
\label{fig:speedofsound}
\end{figure}

Figure \ref{fig:speedofsound} illustrates the evolution of $c_s^2$ as a function of the pressure $P$ for both fluids explained in this section. The conformal limit ($c_s^2 = 1/3$), which characterizes an ultra-relativistic ideal gas ($P = \varepsilon/3$), and the causality limit ($c_s^2 = 1$) are depicted.
The QM EoS and the DM EoS ($n=4$) remain relatively soft, asymptotically approaching the conformal limit from below. In contrast, the DM models for $n=40$ cross the conformal limit at intermediate pressures and rise getting closer to the causality limit ($c_s^2 \to 1$). However, in order to reach the causality limit, the interaction strength must be $n\to\infty$.
The figure also illustrates the effect of the boson mass. For a fixed interaction strength, increasing $m_b$ systematically lowers the speed of sound at any given pressure. Consequently, lighter bosons provide larger pressure support and allow for more massive configurations, whereas heavier bosons lead to lower maximum masses.
These differences in the stiffness of the individual fluids will play a key role in determining the structure of the admixed configurations discussed in the following section.

\section{Results}
\label{sec:results}

This section presents the numerical results obtained for compact stars composed of deconfined QM admixed with self-interacting bosonic DM, using the EoSs described in Section \ref{sec:EoS}. We consider four representative boson masses, $m_b = 200, 300, 500$, and $1000$ MeV, for two interaction regimes, a weakly interacting case ($n=4$) and a strongly interacting case ($n=40$).
For each model, the two-fluid TOV equations are solved over a two-dimensional grid of central pressures $(P_c^{\mathrm{QM}}, P_c^{\mathrm{DM}})$, covering the ranges $P_c^{\mathrm{QM}} \in [10^{-3},10^5]$ MeV/fm$^3$ and $P_c^{\mathrm{DM}} \in [10^{-5},10^7]$ MeV/fm$^3$. These ranges extend from the low-pressure limit of both fluids up to the onset of instability, ensuring that the complete stable sequence, from minimum-mass configurations to the instability turning points, is included. 

The integration of the two-fluid equations provides the mass, radii, and particle numbers of each species. These quantities are then used to perform the stability analysis described in Section \ref{sec:StabAnalysis}. The following sections focus on the stable solutions, analyzing their mass-radius relations, influence of the DM parameters, compactness, and compatibility with current observational constraints.
\subsection{Stability study and compactness} 
\label{sec:results_stability_compactness}

\begin{figure*}[htbp]
    \centering
    \includegraphics[width=0.95\textwidth]{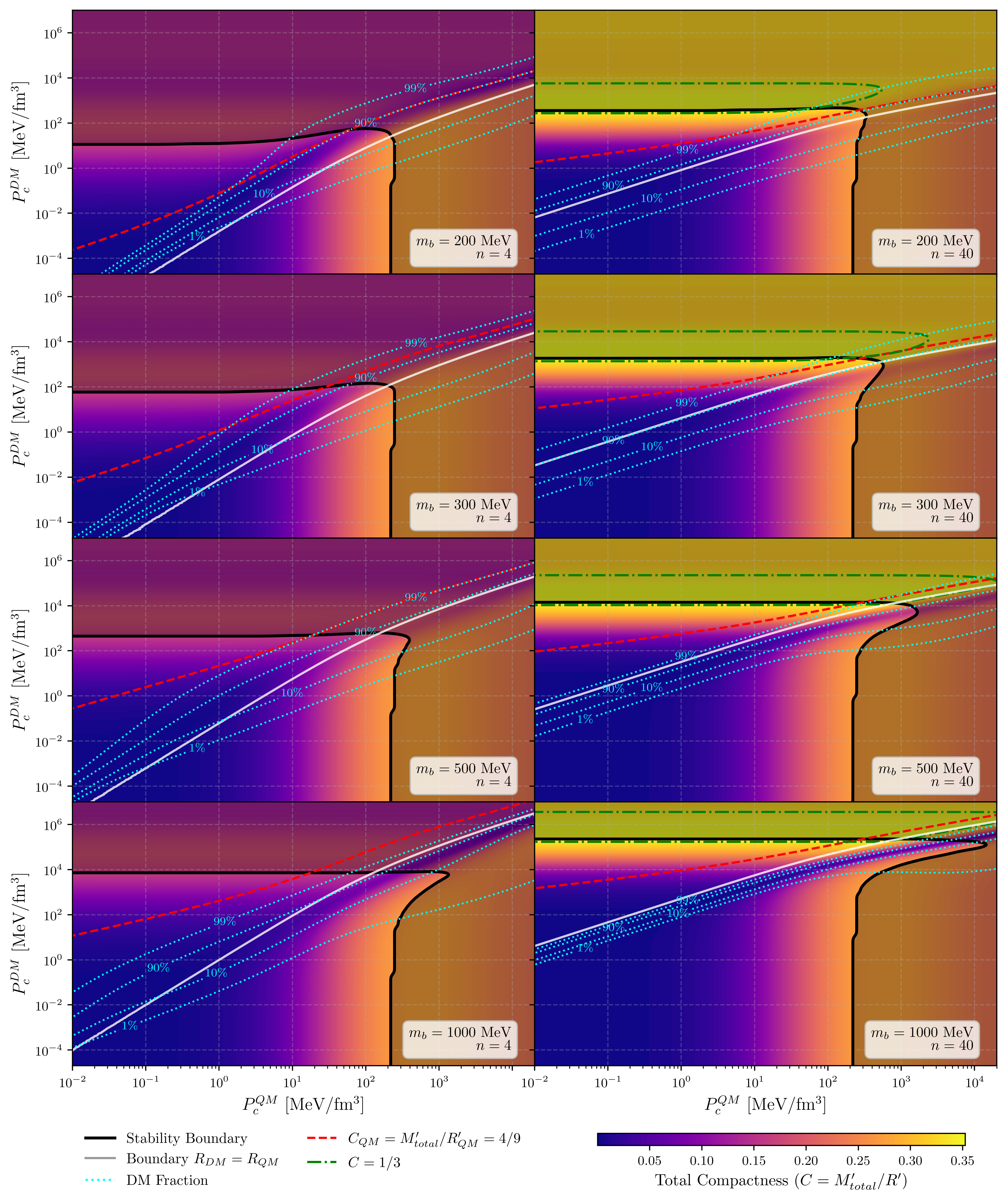}
    \caption{Contour maps of the solution space in the central pressure plane $(P_c^{\mathrm{QM}}, P_c^{\mathrm{DM}})$ for different combinations of the DM boson mass $m_b$ and interaction parameter $n$. The colour scale indicates the total compactness of the star, $C = M'_{\mathrm{total}}/R'$, with $R=max(R_{QM},R_{DM})$. The stability boundary is marked by the solid black line, separating stable and unstable configurations. The white line shows the transition between DM-halos (above the line) and DM-cores (below the line). The dash-dotted cyan lines represent contours of constant DM fraction ($N_{\mathrm{DM}}/N_{\mathrm{total}}$), the dashed green line marks the boundary for ultra-compact stars $C = 1/3$ and the dashed red line indicates the pseudo-compactness Buchdahl limit $C_{\mathrm{QM}} = M'_{\mathrm{total}}/R'_{\mathrm{QM}} = 4/9$.}
        \label{fig:stability_contours}
\end{figure*}

The results of the stability analysis are shown in Figure~\ref{fig:stability_contours}, which displays a grid of contour maps for the distinct combinations of boson mass $m_b$ and interaction parameter $n$ considered in this work.

Each panel in the figure represents the solution space for a specific DM model $(m_b, n)$ mapped onto the grid of central pressures. 
The enclosed region by the solid black contour corresponds to physically stable solutions, while the outside solutions are unstable. The background color scale represents the total compactness, defined as $C = M'_{\mathrm{total}}/R'$ with $R=\max(R_{QM},R_{DM})$ (see Section \ref{subsec:Compactness} for discussion). To characterize the internal structure of the stable configurations, the solid white contour indicates the topological transition where both fluids share the exact same radius ($R_{\mathrm{DM}}=R_{\mathrm{QM}}$), the region above the white line corresponds to configurations where DM forms an extended halo ($R_{\mathrm{DM}} > R_{\mathrm{QM}}$), whereas the region below it represents configurations where DM is confined to a compact inner core ($R_{\mathrm{DM}} < R_{\mathrm{QM}}$). A complete characterization of this structural transition will be provided in Section \ref{subsec:Mass_radii}. Additionally, cyan dotted contours denote constant DM particle fractions. As visually evident, moving upward across the grid toward higher DM central pressures smoothly transitions the system from QM-dominated core configurations to highly DM-dominated halo structures.

Finally, the figure includes two characteristic compactness thresholds. The green contour marks the ultra-compact condition $C=1/3$, corresponding to the formation of a photon sphere, as discussed in Ref.~\cite{Pitz_BosonStars}, while the red dashed contour indicates the {\em apparent} compactness limit of the visible quark component, the pseudo-compactness Buchdahl limit of $C_{\mathrm{QM}}=M'_{\mathrm{total}}/R_{\mathrm{QM}}=4/9$. 
\subsubsection{Dependence of the stability region on the dark matter parameters}

For all the models considered, stable configurations are found at low central pressures of both fluids. Increasing either $P_c^{QM}$ or $P_c^{DM}$ eventually drives the system across the stability boundary, where the star becomes unstable against radial perturbations. The exact shape and extension of this boundary depend strongly on the properties of the DM component.

Comparing the two interaction regimes, for a fixed boson mass, strongly interacting models exhibit larger stable regions than the weakly interacting cases. This behavior is directly associated with the highly incompressible nature of the $n=40$ EoS at high densities. As established in the DM EoS analysis of Section \ref{sec:DM_EoS} and quantitatively demonstrated in Table \ref{tab:dm_max_masses}, strongly interacting DM can sustain central pressures orders of magnitude higher than its weakly interacting counterpart before reaching the critical threshold for gravitational collapse.

The stability maps also show that as $m_b$ increases, the stable region extends toward significantly larger DM central pressures. This larger stable region can be understood through the same mechanism discussed above: increasing the boson mass allows the DM fluid to sustain higher central pressures before reaching instability.

A distinctive feature of the stability maps is the appearance of extended stability regions reaching very high central pressures (see upper right corners of the stability region in all plots). Whether these regions extend to either larger $P_c^{\mathrm{DM}}$ or $P_c^{\mathrm{QM}}$ with $m_b$ and $n$ is related to which component occupies the central region of the star, thus defining the core-halo structure.

For models with heavier bosons or strong self-interactions, the upper right corner of the stable region moves to higher $P_c^{\mathrm{QM}}$. In that region, the stars present a compact DM-core, indicating that it provides the central pressure support needed to maintain the surrounding QM envelope at values of $P_c^{\mathrm{QM}}$ where the pure QM sequence would already become unstable.

A complementary behaviour, but less pronounced, is observed in models with lighter and weakly interacting bosons. In this case, the upper right corner of the stability region extends to high DM central pressures and the compact star configurations have a DM-halo  (above the white line) and thus a QM-core. The self-bound QM-core allows the surrounding diffuse DM-halo to remain in equilibrium at central pressures beyond the instability threshold of a pure bosonic configuration.

These results demonstrate that the stability of admixed compact stars is complex and that a sufficiently stable and compact core can extend the range of central pressures over which the coupled two-fluid configuration remains stable.

\subsubsection{Compactness}
\label{subsec:Compactness}

The total compactness of the stable configurations reaches up to $C \simeq 0.341$ across all parameter sets considered. As shown by the colour scale in Figure \ref{fig:stability_contours}, the maximum compactness is controlled by the DM self-interaction strength and is independent of the boson mass. The $n = 4$ models exhibit a consistent upper bound of $C \simeq 0.269$, which is set by the pure quark star. For $n = 40$, the DM-only branch becomes more compact than the quark star, and every model reaches $C \simeq 0.341$, attained by almost pure DM configurations.

Consequently, only the $n=40$ configurations enter the ultra-compact regime $C \ge 1/3$ (green dashed line), where photon-sphere solutions become possible, as seen by the fact that the green line in Figure \ref{fig:stability_contours} only appears in the right panels. Moreover, the possible configurations occupy a narrow region of parameter space, located between the stability boundary and the ultra-compact threshold, and are restricted to configurations with high DM fractions ($N_{\mathrm{DM}}/N_{\mathrm{tot}} \gtrsim 0.9$) and DM-halos.

A comparison with baryonic-DM models from Pitz et al. \cite{pitz2024_om_dm} shows a qualitative agreement. In both frameworks, ultra-compact configurations are not found in the weakly interacting regime and appear only for strongly interacting DM at high DM fractions and with DM-halos. However, the two cases differ in the region of parameter space where these solutions exist. While baryonic models allow ultra-compact solutions over a relatively broad range of central pressures, the quark-based framework confines them to a much narrower region close to the onset of instability. Nonetheless, it was shown in \cite{Rather:2026aed} that neutron stars with a DM core can also be ultra-compact by including effects from a phase transition in normal matter. Even a small DM fraction is sufficient to increase the compactness to the point where the star develops a light ring outside its surface.

Finally, we compare the models against theoretical compactness limits. The red dashed line marks the pseudo-compactness threshold $C_{QM} = 4/9$, where $ C_{QM} =M_{\mathrm{total}}/R_{QM}.$ As shown in Figure \ref{fig:stability_contours}, in all panels, numerous stable solutions lie above the pseudo-compactness threshold, seemingly violating standard compactness limits when interpreted solely in terms of the QM radius. All such configurations are found to have an extended DM-halo.

Since electromagnetic observations are sensitive only to the visible QM surface, these halo configurations can appear significantly more compact than their true total compactness would suggest. This apparent violation does not indicate physical instability. Although the QM-core may exceed the standard compactness limit in isolation, the overall configuration remains  stable because the extended DM-halo provides additional gravitational support, maintaining equilibrium in the coupled two-fluid system.
\subsection{Mass-radius relations}
\label{subsec:Mass_radii}

\begin{figure*}[htbp]
    \centering
    \includegraphics[width=0.95\textwidth]{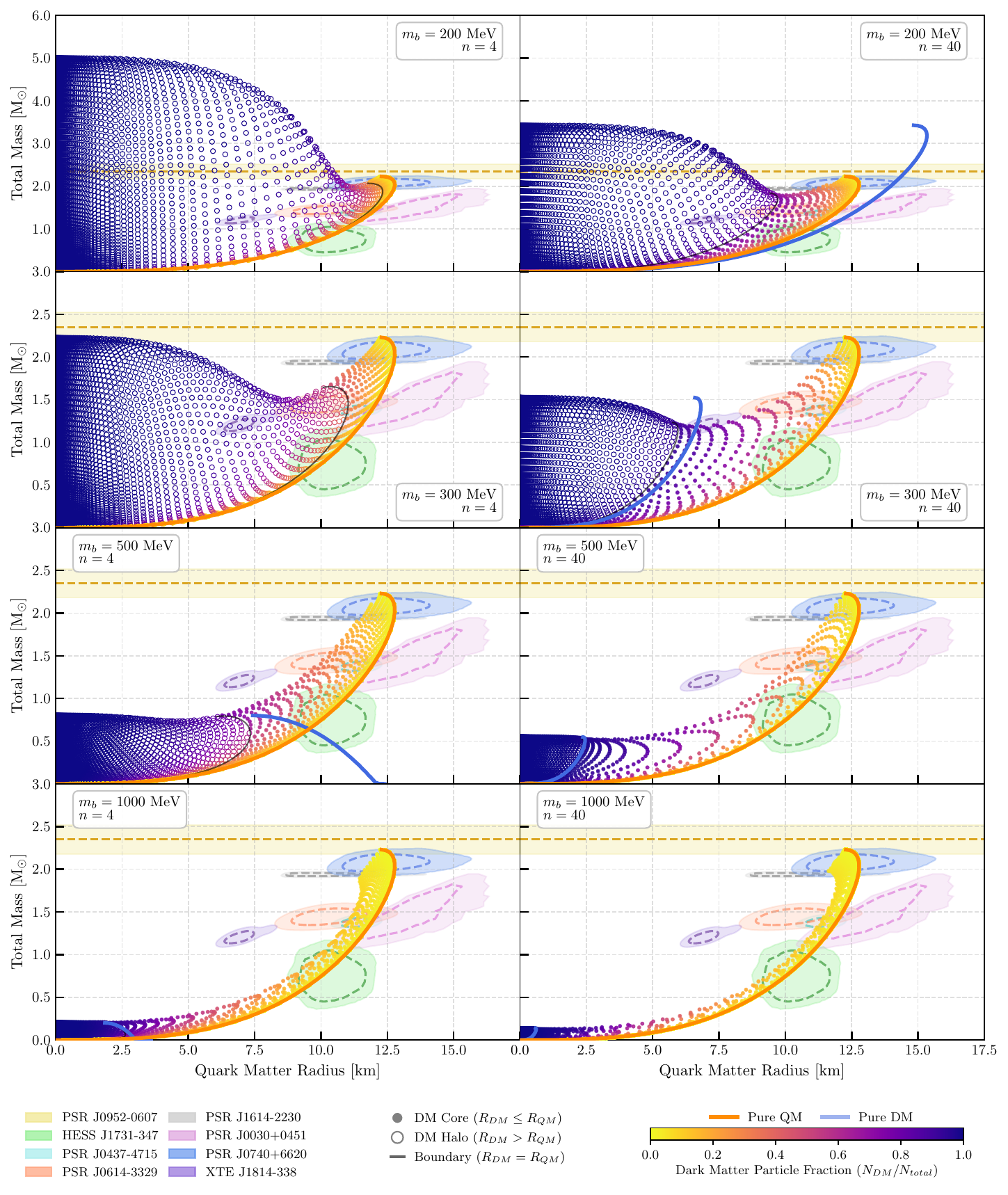}
    
    \caption{Total mass as a function of the QM radius of  stable compact stars for different combinations of the DM boson mass $m_b$ and interaction parameter $n$. The color scale represents the DM particle fraction, while filled and open markers denote DM-core and DM-halo configurations, respectively. The solid black line marks the transition between both. The pure QM sequence (orange) is included in all panels, whereas the pure DM sequence (blue) is shown only when its scale is comparable to that of the admixed solutions.   Shaded contours represent 1$\sigma$ and 2$\sigma$ confidence intervals for the observational measurements of HESS J1731$-$347 \cite{anomaloussmallmass}, PSR J0437$-$4715 \cite{Choudhury_2024}, PSR J0614$-$3329 \cite{Mauviard:2025dmd,Miller:2026vpr}, PSR J1614$-$2230 \cite{Demorest:2010bx,NANOGrav:2017wvv}, PSR J0030$+$0451 \cite{Miller_2019, Riley_2019,Vinciguerra:2023qxq,Kini:2026rjx}, PSR J0740$+$6620 \cite{Riley2021,Miller_2021} and  XTE J1814$-$338 \cite{Kini_2024}. The mass measurement of PSR J0952$-$0607 \cite{MostMassiveNS} is also included.}
    
    \label{fig:final_mr}
\end{figure*}

Figure \ref{fig:final_mr} shows the stable mass-radius configurations obtained from the coupled two-fluid TOV equations. The total gravitational mass is plotted against the visible $R_{QM}$. 
 Shaded contours represent 1$\sigma$ and 2$\sigma$ confidence intervals for the observational measurements of HESS J1731$-$347 \cite{anomaloussmallmass}, PSR J0437$-$4715 \cite{Choudhury_2024}, PSR J0614$-$3329 \cite{Mauviard:2025dmd,Miller:2026vpr}, PSR J1614$-$2230 \cite{Demorest:2010bx,NANOGrav:2017wvv}, PSR J0030$+$0451 \cite{Miller_2019, Riley_2019,Vinciguerra:2023qxq,Kini:2026rjx}, PSR J0740$+$6620 \cite{Riley2021,Miller_2021} and  XTE J1814$-$338 \cite{Kini_2024}. The mass measurement of the 
 black-widow pulsar PSR J0952$-$0607 \cite{MostMassiveNS} is also included.

Following the same structure as in Figure \ref{fig:stability_contours}, the figure is arranged in a grid where each panel corresponds to a specific combination of the DM boson mass $m_b$ and the interaction parameter $n$. The stable admixed star configurations are represented by scatter points and the marker style reflects their characteristics. The color scale represents the DM particle fraction, $N_{\mathrm{DM}}/N_{\mathrm{total}}$.

The marker style distinguishes the structure of the star: filled circles correspond to DM-cores and open circles to DM-halos. The transition between both is marked by the solid black curve ($R_{\mathrm{DM}} = R_{\mathrm{QM}}$).

For reference, the mass-radius relations of the corresponding single-fluid models are also shown. These limiting cases provide the reference boundaries within which the admixed solutions are distributed. The orange line represents the pure QM sequence and it is shown in all the cases, whereas the blue line corresponds to pure DM. However, this blue line is not seen for $n=4$ and $m_b=200,300$ MeV as it lies outside the mass-radius region of the admixed solutions.

The admixed star configurations represent a continuous transition between the two physical boundaries: a pure quark star and a pure boson star. As illustrated by the color gradient in Figure \ref{fig:final_mr}, the solutions smoothly approach the corresponding single-fluid limits, that is the QM or DM baselines, when the central pressure of one of the components becomes negligible.
In the limit where the DM pressure vanishes ($P_c^{\mathrm{DM}} \to 0$), the system is dominated by the quark component (yellow markers). In this regime, the solutions converge toward the solid orange curve, recovering the pure QM mass-radius relation. Conversely, in the limit of vanishing QM central pressure ($P_c^{\mathrm{QM}} \to 0$), the configurations become almost entirely dominated by DM (dark blue markers). 

As shown in the figure, the boson mass $m_b$ and the interaction strength $n$ strongly affect the DM-dominated configurations (dark blue markers). The maximum mass of an admixed configuration is limited by the upper limit of either the pure QM or DM EoS. The DM parameters only modify the maximum achievable mass when the pure DM sequence exceeds the maximum mass supported by pure QM. Otherwise, the pure QM EoS sets the upper mass limit. 
The highest mass is achieved with the smallest value of $m_b=200$~MeV. Lowering this mass even more would result in even larger configurations and thus would lie outside of the M-R range for pure neutron or quark stars. 

Also, for a fixed parameter set (e.g., $m_b=200$ MeV, $n=4$), one can observe a transition between core and halo configurations. It is mainly governed by the dominance of one fluid over the other, that is, the component with the larger central pressure and energy density tends to dominate the inner gravitational potential, forcing the other component to remain confined within it. This is illustrated in Figure \ref{fig:pressure_profiles} of Appendix~\ref{appendix}. 

As for the extension of the DM-halos, we find that the weakly interacting models produce the most extensive halos, particularly for light bosons, with extensions reaching $R_{DM} - R_{QM} \sim 75$ km for $m_b=200$ MeV. As the boson mass and the self-interaction increase, the DM extension strongly decreases.

\subsection{Observational constraints}
\label{subsec:Observational_constraints}

In order to determine the possible constraints on the DM parameters coming from observations, in Fig.~\ref{fig:final_mr} we compare the predicted mass-radius relations with several astrophysical measurements. Specifically, we consider the constraints on the mass-radius relation from the observation of HESS J1731$-$347 \cite{anomaloussmallmass}, PSR J0437$-$4715 \cite{Choudhury_2024}, PSR J0614$-$3329 \cite{Mauviard:2025dmd,Miller:2026vpr}, PSR J1614$-$2230 \cite{Demorest:2010bx,NANOGrav:2017wvv}, PSR J0030$+$0451 \cite{Miller_2019, Riley_2019,Vinciguerra:2023qxq,Kini:2026rjx}, PSR J0740$+$6620 \cite{Riley2021,Miller_2021}, and XTE J1814$-$338 \cite{Kini_2024}, together with the mass measurement of PSR J0952$-$0607 \cite{MostMassiveNS}. 

As seen in Fig.~\ref{fig:final_mr}, all these astrophysical measurements, except for XTE J1814$-$338, can be reproduced for the explored parameter space, favoring compact star configurations with low DM content. As for the XTE J1814$-$338 case, only the parameter sets with $m_b=200$ MeV and $m_b=300$ MeV are capable of reproducing it for both DM interaction strengths. 

\begin{figure*}[!ht]
    \centering
    \includegraphics[width=\linewidth]{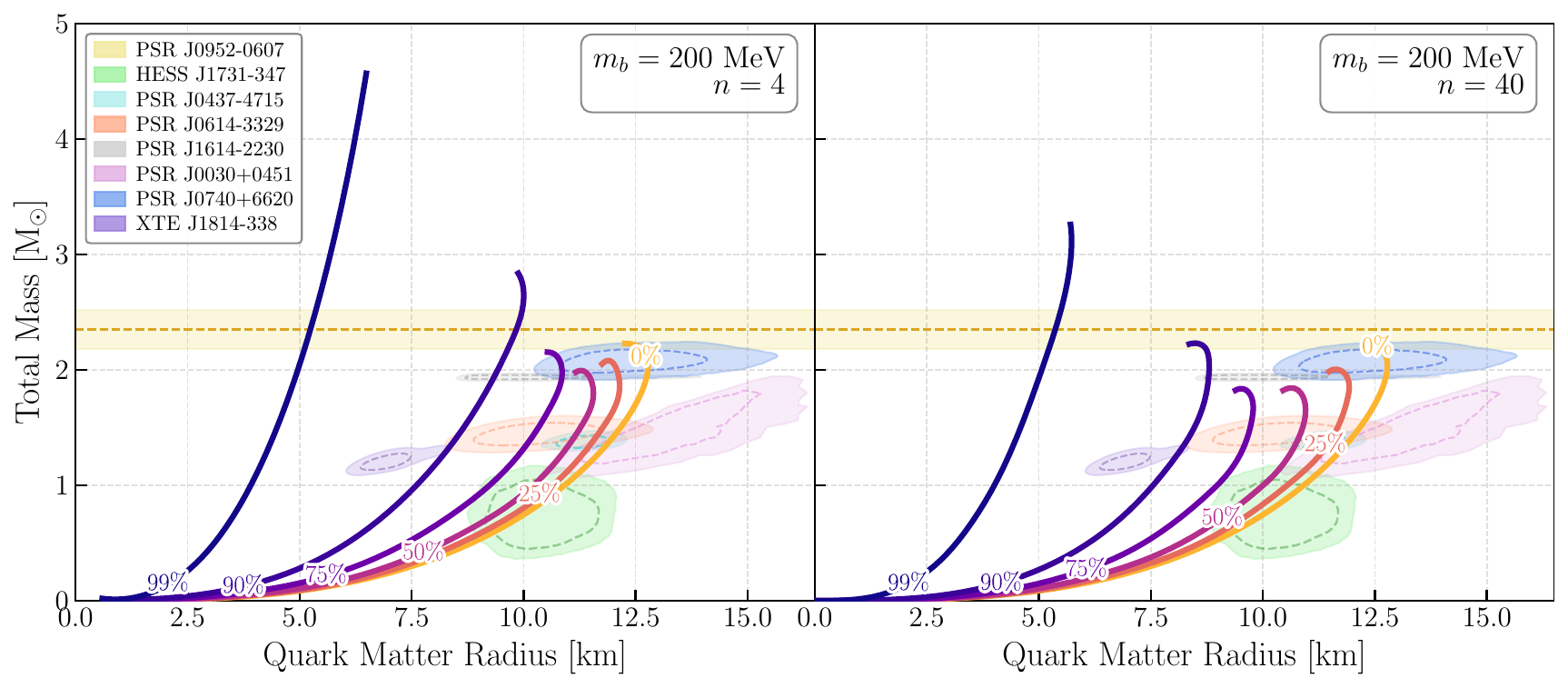}
    \caption{Total mass and radius $R_{QM}$ for fixed total DM particle fractions. The left panel shows the results for $m_b=200$ MeV and $n=4$, while the right panel displays the $m_b=200$ MeV and $n=40$ case. Shaded contours represent 1$\sigma$ and 2$\sigma$ confidence intervals for the observations of HESS J1731$-$347 \cite{anomaloussmallmass}, PSR J0437$-$4715 \cite{Choudhury_2024}, PSR J0614$-$3329 \cite{Mauviard:2025dmd,Miller:2026vpr}, PSR J1614$-$2230 \cite{Demorest:2010bx,NANOGrav:2017wvv}, PSR J0030$+$0451 \cite{Miller_2019, Riley_2019,Vinciguerra:2023qxq,Kini:2026rjx}, PSR J0740$+$6620 \cite{Riley2021,Miller_2021} and  XTE J1814$-$338 \cite{Kini_2024}. The mass measurement of the PSR J0952$-$0607 \cite{MostMassiveNS} is also included.}
    \label{fig:TotalMass_vs_QRadius_Dual_Plot.png}
\end{figure*}

To illustrate these light-boson scenarios, Figure~\ref{fig:TotalMass_vs_QRadius_Dual_Plot.png} shows the mass-radius relations for the cases $m_b=200$ MeV with $n=4,40$ along with the observational constraints. Except for PSR J1614$-$2230 and XTE J1814$-$338, all other constraints can be already fulfilled with nearly pure QM configurations. As we increase the DM amount it becomes more difficult to satisfy all observational constraints at the same time. In fact, compact star configurations with a DM fraction above 75\% do not satisfy the mass and radius constraints from HESS J1731$-$347, PSR J0437$-$4715, PSR J0030$+$0451, and PSR J0740$+$6620. 
As for the PSR J1614$-$2230 mass-radius constraint, we find that in order to fulfill this requirement the amount of DM should be between 25\% and 90\%, the specific value depending on the values of $m_b$ and $n$. 
For XTE J1814$-$338, the observed combination of mass and radius is difficult to reproduce with pure QM which is also the case for standard purely baryonic models, see Ref.~\cite{pitz2024_om_dm}. In our admixed configurations, matching the observed mass-radius combination requires a substantial DM component, with more than 90\%.
We also note that the PSR J0952$-$0607 mass constraint can be obtained with either QM-dominated configurations or DM-dominated, but in the latter case only for extremely low radii.

In summary, the observational constraints favour the parameters $m_b=200,300$ MeV for both $n=4$ and $n=40$. These models allow for extended DM-halos in the most extreme cases and the formation of DM-dominated configurations. 

\begin{figure*}[th!]
    \centering
    \includegraphics[width=\linewidth]{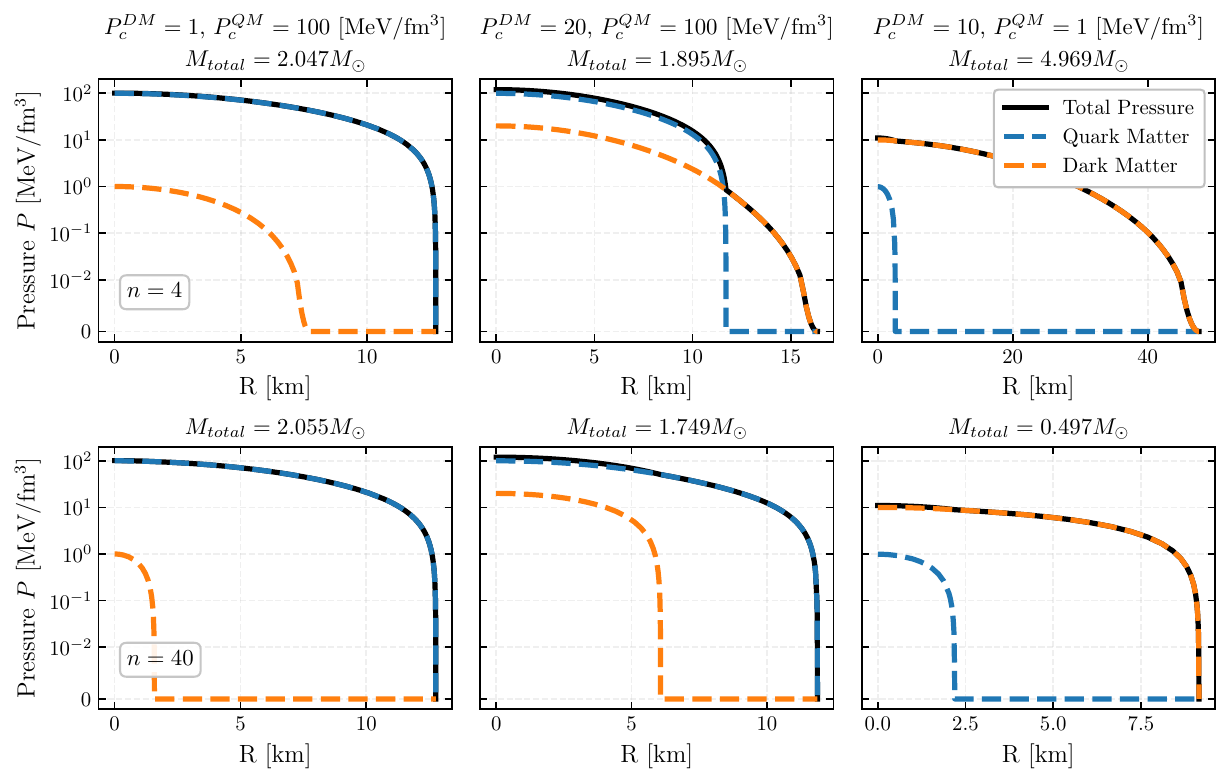}   
    \caption{Radial pressure profiles for representative two-fluid compact star configurations. The top row illustrates configurations for a weakly interacting DM model ($m_b = 200$ MeV, $n=4$), while the bottom row corresponds to a strongly interacting DM model ($m_b = 200$ MeV, $n=40$). Each column corresponds to a specific pair of central pressures $(P_c^{\mathrm{DM}}, P_c^{\mathrm{QM}})$. The dashed blue and dashed orange lines show the partial pressures of the QM and DM components, respectively, while the solid black line represents the total pressure of the system. The radial point where each partial pressure drops to zero defines the physical surface ($R_{\mathrm{QM}}$ or $R_{\mathrm{DM}}$) of that component.}
\label{fig:pressure_profiles}
\end{figure*}

\section{Conclusions}
\label{sec:conclusions}

In this work, we have investigated the stability, compactness, and mass-radius relations of compact stars composed of QM admixed with self-interacting bosonic DM. By exploring a comprehensive grid of central pressures, we have mapped the stellar properties across different boson masses $m_b$ and self-interaction strengths $n$.

The stability analysis reveals that both the interaction strength and the boson mass critically govern the viable parameter space. Strongly interacting DM models ($n=40$) exhibit significantly larger stable regions than their weakly interacting counterparts ($n=4$), as the highly incompressible limit of the strongly interacting DM fluid provides robust structural support. Similarly, increasing the boson mass allows the DM fluid to sustain higher central pressures before reaching instability. Altogether, the interplay of both fluids extends the range of stable configurations beyond that of pure stars.

This stabilizing effect is also closely connected to the emergence of ultra-compact configurations. Within the parameter space explored in this work, stable ultra-compact solutions ($C \ge 1/3$) are found only for strongly interacting DM and very large DM fractions ($N_{DM}/N_{tot} \gtrsim 90\%$). These configurations consist of a highly compressed QM core surrounded by a DM-halo and are compact enough to develop a photon sphere, making them potential black-hole mimickers. However, unlike true black holes, they do not possess an event horizon. Consequently, the possible observation of a photon ring together with electromagnetic emission from a visible surface could provide a distinctive signature of these ultra-compact DM-admixed stars.

Compared to previous studies based on baryonic matter backgrounds, ultra-compact solutions are considerably more restricted in the QM-admixed scenario. The softer QM EoS imposes much more stringent conditions on the DM sector, confining ultra-compact stars to a small region of the parameter space nearly suppressing their existence.

A notable result is the distinction between the true compactness of the star and the compactness inferred from electromagnetic observations. In halo configurations, the visible radius corresponds only to the QM component, while a substantial fraction of the total mass is distributed in an extended DM-halo. As a result, the compactness inferred from the visible radius can greatly exceed the value expected for ordinary compact stars. This provides a potential observational signature of DM-admixed compact stars also for the QM case.

The mass-radius relations of these admixed configurations are fundamentally bounded by the single-fluid limits of the pure EoS. When the pure QM EoS sets the absolute upper mass limit, the maximum mass of the admixed configurations is strictly capped. Conversely, when the pure DM sequence dictates the upper boundary, DM-dominated stars can achieve exceptionally high masses. We observe that increasing $m_b$ or $n$ reduces the maximum achievable masses as the contribution of DM becomes progressively suppressed and the solutions approach the pure QM sequence.

Structurally, the core-halo structure is governed by the relative boundary conditions of both fluids. Weakly interacting DM forms massive, extended halos, whereas strongly interacting DM mimics a self-bound fluid, favoring compact DM cores. Likewise, increasing $m_b$ suppresses the spatial extension of the DM component, favoring compact cores and QM halos.
Comparison against observational constraints reveals that PSR J1614$-$2230 and XTE J1814$-$338 may serve as indicators for the presence of DM. Specifically, XTE J1814$-$338 might offer a key probe into the dark sector, requiring light boson masses ($m_b \approx 200\text{--}300\text{ MeV}$) alongside heavily DM-dominated configurations.
Crucially, the configurations capable of satisfying the XTE J1814$-$338 constraint is associated with extended DM halos. However one needs to be cautious with applying NICER observations to the case of a neutron star with a DM halo. The surface of the neutron star where the emitting hotspots are located are embedded in a DM environment. This gives an additional gravitational redshift due to the stronger gravitational potential. This redshift is then dictated by the timelike component of the metric tensor $g_{tt}$, given by $\alpha$ in \ref{eq:alpha}. The values of the redshift differ significantly from the case where the surface is matched to the vacuum exterior (i.e.\ where the gravitational radius lies), causing a seemingly violation of the redshift Buchdahl limit of $z = 2$ \cite{Rather:2026aed}. 
Moreover we note that extended DM halos may have important consequences for multi-messenger astronomy, as the increased stellar size can significantly modify the tidal deformability measured in gravitational-wave observations of binary mergers, providing an independent probe of the presence of DM in compact stars.
The parameter space explored in this work is limited to a specific range of boson masses and self-interaction strengths. Further studies should explore a wider range of parameters, including lower boson masses and higher self-interaction strengths, to fully map the landscape of DM-admixed quark stars.

\section*{Acknowledgments}
J.T.B. acknowledges support from CSIC under JAE Intro ICU 2025 program (No.~JAEICU$\_$25$\_$03185).   S.L.P. and J.S.B. acknowledge support by the Deutsche Forschungsgemeinschaft (DFG, German Research Foundation) through the CRC-TR 211 ’Strong-interaction matter under extreme conditions’– project number 315477589 – TRR 211. S.L.P. is supported by the Research and Development (R$\&$D) program of GSI Helmholtzzentrum für Schwerionenforschung. L.T. acknowledges support by the projects PID2022-139427NB-I00 and PID2025-168786NB-I00 financed by the Spanish MCIN/AEI/10.13039/501100011033 and ERDF/EU, and by the Grant CIPROM 2023/59 of Generalitat Valenciana.

\appendix
\section{Appendix}
\label{appendix}

In this Appendix we discuss the radial profiles for few representative two-fluid compact star configurations.

Figure \ref{fig:pressure_profiles} shows the radial pressure profiles of both fluids for three representative central pressure pairs $(P_c^{\mathrm{DM}}, P_c^{\mathrm{QM}})$. The three panels demonstrate how the partial pressures of the QM (dashed blue lines) and DM (dashed orange lines) evolve from the center of the star to its boundary for $n=4$ (top panels) and $n=40$ (bottom panels). The total pressure, which supports the star against gravitational collapse, is represented by the solid black line.

When QM dominates the central region (left panel), the DM remains confined inside the visible quark surface, resulting in a DM-core, which corresponds to the high-QM-fraction configurations (yellow filled markers) in Figure \ref{fig:final_mr}.
As the DM contribution increases (middle panel), the DM pressure profile extends outward and approaches the transition between both topologies $R_{\mathrm{DM}} = R_{\mathrm{QM}}$. For this specific pair of pressures we can see that the $n=4$ case has a DM-halo while the $n=40$ case has a DM-core. In Figure \ref{fig:final_mr}, solutions with radial profiles resembling this panel are located close to the topological boundary (solid black line) separating filled and hollow markers in the mass-radius relations.

Finally, for DM-dominated solutions (right panel), the QM is compressed into a compact central object forming a DM-halo.
Comparing the top and bottom rows of Figure \ref{fig:pressure_profiles}, we observe that strongly interacting models exhibit a more abrupt drop in the DM pressure profile, leading to a highly compact DM distribution. This distinct spatial behaviour originates directly from the low-pressure characteristics of the DM EoS. While the central QM is inherently self-bound, weakly interacting DM behaves as a diffuse, gravity-bound fluid whose energy density gradually decreases outward. Once the edge of the QM core is reached, this smooth decline enables the weakly interacting DM to naturally expand and form massive extended halos (reaching 47 km in radius while the QM core is reduced to 2.6 km in the right panels of Figure \ref{fig:pressure_profiles} as denoted by the dark blue hollow markers in Figure \ref{fig:final_mr}). Conversely, strongly interacting DM maintains a finite energy density even at remarkably low pressures, effectively mimicking a self-bound object. Consequently, it cannot form a diffuse envelope and remains compactly concentrated as a mantle around the QM core.

\bibliography{apssamp}
\bibliographystyle{apsrev4-1}

\end{document}